\documentclass[hyper]{prop2015}% no class options needed by now
\usepackage[english]{babel}

\usepackage{tikz}
\usepackage[all]{xy}

\usepackage{amsthm}

\usepackage{bbm}

\theoremstyle{definition}
\newtheorem{definition}{Definition}[section]
\newtheorem{informaldefinition}[definition]{Informal definition}

\newtheorem{example}[definition]{Example}

\theoremstyle{plain}

\newtheorem{proposition}[definition]{Proposition}
\newtheorem{lemma}[definition]{Lemma}

\def\tilde{\sim}
\def\oone{\mathbb{1}}
\def\ops{ { \{ * \} } }

\def\mR{\mathbb{R}}
\def\mL{\mathbb{L}}

\def\AA{\mathfrak{A}}
\def\CC{\mathfrak{C}}
\def\DD{\mathfrak{D}}
\def\QQQ{\mathfrak{Q}}
\def\UU{\mathfrak{U}}
\def\VV{\mathfrak{V}}

\def\OO{\mathcal{O}}
\def\PP{\mathcal{P}}
\def\QQ{\mathcal{Q}}
\def\RR{\mathcal{R}}

\def\CCC{\mathbf{C}}
\def\DDD{\mathbf{D}}
\def\EEE{\mathbf{E}}
\def\MMM{\mathbf{M}}

\def\Alg{\mathbf{Alg}}
\def\Set{\mathbf{Set}}
\def\Vec{\mathbf{Vec}}
\def\Ch{\mathbf{Ch}}
\def\Symp{\mathbf{Symp}}
\def\Loc{\mathbf{Loc}}
\def\COpens{\mathbf{COpens}}
\def\Seq{\mathbf{Seq}}
\def\SymSeq{\mathbf{SymSeq}}
\def\Op{\mathbf{Op}}
\def\FT{\mathbf{FT}}
\def\QFT{\mathbf{QFT}}
\def\CFT{\mathbf{ClFT}}
\def\LFT{\mathbf{LFT}}
\def\OrthCat{\mathbf{OrthCat}}

\def\As{\mathrm{As}}
\def\Lie{\mathrm{Lie}}
\def\uLie{\mathrm{uLie}}
\def\Pois{\mathrm{Pois}}
\def\Diag{\mathrm{Diag}}
\def\MDiag{\mathrm{MDiag}}

\def\CCR{\mathfrak{CCR}}

\def\id{\text{id}}
\def\Mor{\text{Mor}}
\def\Heis{\text{Heis}}

\newcommand\und[1]{\underline{#1}}
\newcommand\ovr[1]{\overline{#1}}
\newcommand\ipt[2]{ \big( \substack{ {#1 }\\ {#2} } \big) }

\category{Proceedings}


\subtitle{\href{http://www.maths.dur.ac.uk/lms/109/index.html}{LMS/EPSRC Durham Symposium on Higher Structures in M-Theory}}

\title{Coloring Operads for Algebraic Field Theory}

\author[S. Bruinsma]{Simen Bruinsma\inst{a,}\footnote{Corresponding author e-mail:~\href{mailto:simen.bruinsma@nottingham.ac.uk}{\textsf{simen.bruinsma@notting\-ham.ac.uk}}}}
\address[1]{School of Mathematical Sciences, University of Nottingham, University Park, Nottingham, NG7 2RD, United Kingdom}

\begin{acknowledgements}
These proceedings are based on joint work with Alexander Schenkel, \cite{Bruinsma:2018knq}. 
I would like to thank him, Marco Benini and Lukas Woike 
for the discussions that led to this work. 
I am supported by a PhD scholarship of the Royal Society (UK). 
\end{acknowledgements}

\begin{abstract}
In these proceedings we summarize previous work where we formalize a general concept of algebraic field theories using operads. After giving a gentle reminder of algebraic quantum field theory, operads and their algebras, we construct field theory operads, whose algebras are exactly algebraic field theories. Specifically, they satisfy a suitable version of the Einstein causality axiom. From this construction we get adjunctions between different types of field theories, including adjunctions related to local-to-global extensions and the time-slice axiom, and a quantization functor for linear field theories that is compatible with these structures. We also take first steps towards a derived linear quantization functor.  
\end{abstract}
\shortabstract

\begin{document}
\maketitle

%%% Use this if the article text won't start with a \section:
% \noindent

\section{Introduction}

Algebraic quantum field theory (AQFT) is a mathematical framework 
to define and study quantum field theories on Lorentzian space-times. 
At its core, an algebraic quantum field theory takes a space-time and all its (causally convex) subspace-times, 
and for each it defines an (associative) algebra of observables in a consistent way. 
This consistency means that an AQFT is a functor 
from a space-time category to an algebraic category. 
Moreover, the assignment of algebras is required to satisfy two physically motivated conditions,
Einstein causality and the time-slice axiom. 

This article describes a way of formalizing the structures found in AQFT using the theory of colored operads. 
Operads are algebraic objects that capture the structure of algebras. 
The crucial step is to construct an operad that encodes the Einstein causality property. 
To do this, we follow the strategy set out in \cite{Benini:2017fnn}. 
We generalize the operads constructed therein, 
allowing for more general field theories that are described by other algebras. 
The results have earlier appeared in \cite{Bruinsma:2018knq}, 
where we will refer to for the proofs that we omit. 

The upshot of this abstract approach is that it provides a framework to formalize universal constructions. 
For example, from this vantage point it is immediate that the category of quantum field theories is bicomplete: 
it contains all limits and colimits. 
Moreover, from maps between operads or space-time categories
we get adjunctions between the related field theories, 
which allow us to study properties like the time-slice axiom and descent.
Lastly, this approach is most suited for moving to higher categorical settings, 
which we need for gauge theory. 

The rest of the article is organized as follows: 
in Section \ref{prelims} we introduce the two main ingredients, 
algebraic quantum field theory and colored operads with their algebras. 
In Section \ref{operadsinFT} we construct our central object of study, 
the colored operad $\PP_{\ovr{\CCC}}^r$ whose algebras are field theories. 
In Section \ref{adjunctions} we describe some general adjunctions
that we get from natural choices of functors,
expressing local-to-global extensions and time-slicification. 
We also construct a specific adjunction which describes the quantization of linear theories. 
In Section \ref{homotopy} we take our first steps towards a homotopy treatment of the same, 
which will hopefully lead to examples of gauge theories in algebraic quantum field theory. 
Lastly, in Section \ref{outlook} we give a short review of remaining open questions.

\section{Preliminaries}\label{prelims}

\subsection{Algebraic quantum field theory}\label{AQFT}

In algebraic quantum field theory \cite{Haag:1963dh,Haag:1992hx,Brunetti:2001dx}, 
we assign to each space-time an algebra of observables. 
The underlying basic idea 
is that locality allows us to also define such an algebra
on any subspace-time of the space-time we are interested in. 
This assignment has to be consistent with space-time embeddings, 
and needs to satisfy certain physically motivated conditions. 

Before we give a definition, we introduce some terminology related to Lorentzian geometry, 
referring to \cite{Benini:2015bsa} for more details. 
A {\it causal curve} in a space-time is a curve that is everywhere lightlike or timelike. 
An open subset $N$ of a space-time $M$ is called {\it causally convex}
if every causal curve in $M$
with endpoints in $N$ wholly lies in $N$. 
We write $\COpens(M)$ for the category of causally convex open subsets of $M$, 
with inclusions $\iota_{N_1}^{N_2} : N_1 \subseteq N_2$ as morphisms. 
Two subsets of a space-time are {\it causally disjoint}
or {\it spacelike separated} if there exists no causal curve
from one of the space-times to the other. 
A codimension 1 hypersurface $\Sigma $ of a space-time $M$ is called {\it Cauchy}
if each maximally extended timelike curve in $M$ intersects $\Sigma$ exactly once; 
we think of $\Sigma$ as a spatial slice of $M$. 
A {\it globally hyperbolic} space-time $M$ 
is a space-time that contains a Cauchy surface $\Sigma$. 

\begin{definition}
An {\it algebraic quantum field theory} on a globally hyperbolic space-time $M$
is a functor 
\begin{subequations}
\begin{equation}
\AA: \COpens(M) \to \Alg
\end{equation}
from the category of causally convex subsets of $M$
to the category of associative algebras
\footnote{In these proceedings we will not consider $*$-structures, 
or more generally involutive categories. 
For a treatment of these in the operadic framework we refer to \cite{Benini:2018mcq}}. 
It is required to satisfy: 
\begin{enumerate}[i)]
\item {\it Einstein causality}: 
if $N_1 , N_2 \subseteq N $ are two causally disjoint subsets of $N \subseteq M$, 
the images of their algebras commute: 
\begin{equation}
 \Big[ \AA \big( \iota_{N_1}^{N} \big) \big( \AA (N_1) \big) , \AA \big( \iota_{N_2}^{N} \big) \big( \AA (N_2) \big) \Big] = \{0\} \subseteq \AA (N)~; 
 \end{equation}
\item {\it the time-slice axiom}: if $\iota_{N_1}^{N_2} : N_1 \subseteq N_2$ is an inclusion 
such that $N_1$ contains a Cauchy surface of $N_2$, 
\begin{equation}
 \AA \big( \iota_{N_1}^{N_2} \big) : \AA (N_1) \stackrel{\cong}{\longrightarrow} \AA (N_2)
 \end{equation}
is an isomorphism. 
\end{enumerate}
\end{subequations}
\end{definition}

Einstein causality makes sure the theory is causal: 
observables on two spacelike separated subsets will not interact. 
With the time-slice axiom, the theory has a sense of dynamics: 
the algebra on a neighborhood of a spacelike surface
determines the algebras in its domain of dependence. 

One way of extending this definition is by considering a more general space-time category than $\COpens(M)$. 
One usually works on the category $\Loc$ of all globally hyperbolic space-times of dimension $n$, 
with isometric embeddings with causally convex image as morphisms
(or rather, with a small category that is equivalent with $\Loc$). 

More generally, one can work with any small category $\CCC$ 
with an {\it orthogonality relation} $\perp \subseteq \{ (f_1: c_1 \to t, f_2: c_2 \to t) \} $ on $\CCC$: 
a set of pairs of maps in $\CCC$ with the same target,
that is symmetric (if $(f_1,f_2) \in \perp$, $(f_2,f_1) \in \perp$) and closed under pre- and post-composition. 
We call $\overline{\CCC} = (\CCC, \perp)$ an {\it orthogonal category} 
and we write $f_1 \perp f_2$ if $(f_1,f_2) \in \perp$. 
In $\Loc$, $\perp$ would be the set of pairs of maps $(f_1: N_1 \to N, f_2: N_2 \to N)$
such that $f_1(N_1)$ and $f_2(N_2)$ are causally disjoint in $N$. 
A morphism of orthogonal categories $F: (\CCC , \perp_\CCC) \to (\DDD , \perp_\DDD)$
is a functor $F: \CCC \to \DDD$ 
such that $F ( \perp_\CCC) \subseteq \perp_\DDD$. 
We write $\OrthCat$ for the category of orthogonal categories. 
For the time-slice axiom we also need a distinguished set of Cauchy morphisms in $\CCC$, which we call $W$. 

Another way to generalize the definition is to consider more general algebras: 
for example, classical field theories have Poisson algebras of observables,  
while for linear field theories we have Heisenberg Lie algebras of linear observables
(this is expanded on in Section \ref{linqu}). 

This leads to the following informal definition, 
which we make precise later. 

\begin{informaldefinition}\label{defFT1}
A {\it field theory} is a functor 
\begin{equation}
 \AA : (\overline{\CCC}, W) \to \mathbf{A} 
 \end{equation}
from an orthogonal category with a set of Cauchy morphisms
to an algebraic category
that satisfies: 
\begin{enumerate}[i)]
\item Einstein causality: if $(f_1: c_1 \to t) \perp (f_2: c_2 \to t)$, 
$\AA(f_1) \big( \AA(c_1) \big)$ and $\AA(f_2) \big( \AA(c_2) \big)$ ``commute'' in $\AA(t)$; 
\item the time-slice axiom: if $f \in W$, $\AA(f)$ is an isomorphism. 
\end{enumerate}
\end{informaldefinition}

Note that we have cheated twice in this definition: 
we have not defined what an algebraic category is, 
and as such we do not know what commuting is in this category. 
We will be able to give a real definition after the next sections. 

It is not at all clear that the category of field theories is a nice category
when defined in this way, as a subcategory of functors satisfying certain conditions. 
To remedy this, we will rewrite our definition to be more natural
and more suited for universal constructions
and constructions in higher categorical settings (i.e. gauge theory settings). 
The time-slice axiom is implemented in a relatively straightforward way: 
functors $\AA: (\CCC, W) \to \mathbf{A}$ that satisfy it are in one-to-one correspondence 
with functors $\AA : \CCC [ W^{-1} ] \to \mathbf{A}$, 
where $\CCC [ W^{-1} ] $ is the localization of $\CCC$ at $W$. 
Formalizing Einstein causality is more involved; 
for this, we work with operads. 

\subsection{Operads}\label{operads}

Operads are structures that encode the operations on algebras. 
Before we define them, we introduce some notation and underlying concepts. 

For any category $\CCC$, we denote by $\CCC (c_1, c_2)$ 
the set of morphisms in $\CCC$ from $c_1$ to $c_2$, 
and for a small category $\CCC$, we write $\CCC_0$ for its set of objects.
We write $\DDD^\CCC$ for the category of functors from $\CCC$ to $\DDD$. 

A {\it monoidal category} $\MMM$ is a category 
with a tensor product $\otimes: \MMM \times \MMM \to \MMM$ 
and a unit object $I \in \MMM$, 
together with natural isomorphisms 
$(m_1 \otimes m_2 ) \otimes m_3 \cong m_1 \otimes (m_2 \otimes m_3)$ for all $m_i \in \MMM$
and $m \otimes I \cong m \cong I \otimes m$ for any $m \in \MMM$. 
$\MMM$ is {\it symmetric} if it has a braiding $\tau: m_1 \otimes m_2 \xrightarrow{\tilde} m_2 \otimes m_1$
such that $\tau^2 \cong \id$, 
and it is called {\it closed} if $(-) \otimes m$ has a right adjoint $[m,(-)]$, 
i.e. $\MMM (m_1 \otimes m , m_2) \cong \MMM (m_1 , [m,m_2])$ naturally.

We will work with a fixed closed symmetric monoidal category $\MMM$ 
and we further assume $\MMM$ to be bicomplete (it contains all limits and colimits). 
A remark on notation: 
for clarity we will assume that $\MMM$ is a concrete category,
so we can work with elements of objects. 
For a coproduct 
\begin{equation}
m \xrightarrow{\quad \iota_s \quad} \coprod_{s \in S} m
\end{equation}
we will denote the elements $ \iota_s (x)$ by $(s,x)$
where $s \in S$, $x \in m$, 
and we will write $[x]$ for elements in the coequalizer $m$ of  
\begin{equation}
\xymatrix@C=2.5em
{ 
m_1
\ar@<0.5ex>[r]^-{r_1}  
\ar@<-0.5ex>[r]_-{r_2}
~&~
m_2
\ar@{-->}[r]  
~&~  
m
}
\end{equation}
where $x \in m_2$. 

Examples of concrete bicomplete closed symmetric monoidal categories are 
$\Set$, the category of sets with Cartesian product and $I = \ops$ the one-point set,  
$ \Vec_k$, the category of vector spaces over a field $k$ with the regular tensor product and $I = k$, 
and $ \Ch(k)$, the category of chain complexes of vector spaces over $k$ 
with the usual tensor product and $I$ the complex with only $k$ in degree 0.

For a set $\CC$ we have $\CC^{n+1}$, 
elements of which we write as $\big( \substack{t \\ \und{c}} \big) $ 
where $t \in \CC$ and $\und{c} = (c_1, \dots c_n) \in \CC^n$. 
We write $|\und{c}| = n$ for the length of $\und{c} = (c_1, \dots c_n)$
and we will call $\CC$ a set of colors. 
A {\it sequence} $X$ on $\CC$ in $\MMM$ is an assignment
$\big( \substack{t \\ \und{c}} \big) \mapsto X \big( \substack{t \\ \und{c}} \big) \in \MMM$
for all $\big( \substack{t \\ \und{c}} \big) \in \CC^{n+1}$ and all $n \geq 0$. 
Equivalently, if we view $\coprod_{n\geq 0} \CC^{n+1}$ as a discrete category 
(a category with only identity morphisms),
a sequence is a functor $X: \coprod_{n\geq 0} \CC^{n+1} \to \MMM$. 
A morphism between two sequences $\phi: X \to Y$ is then a natural transformation between these functors, 
i.e. a family of maps $X \ipt{t}{\und{c}} \to Y \ipt{t}{\und{c}}$. 
We write $\Seq_\CC(\MMM)$ for the category of sequences on $\CC$ in $\MMM$.

A {\it symmetric sequence} $X$ on $\CC$ in $\MMM$ 
is a sequence on $\CC$ in $\MMM$ with a right-action of the symmetric group: 
for $\sigma \in \Sigma_n$ a permutation on $\{1, \dots, n \}$ 
and $\und{c} = (c_1, \dots c_n )$, 
\begin{equation}
 X_\sigma: X \big( \substack{t \\ \und{c}} \big) \to X \big( \substack{t \\ \und{c} \sigma} \big)~.
 \end{equation} 
Here, $\und{c} \sigma = (c_{\sigma(1)}, \dots, c_{\sigma(n)} ) $. 
If we define $\Sigma_\CC$ to be the groupoid 
with elements of $\coprod_{n\geq 0} \CC^{n+1}$ as objects 
and morphisms given by the action of $\Sigma_n$, 
a symmetric sequence is equivalently a functor $X: \Sigma_\CC \to \MMM$. 
A morphism between two symmetric sequences is a natural transformation between the two functors, 
i.e. a family of maps that respects the symmetric action. 
We write $\SymSeq_\CC(\MMM)$ for the category of symmetric sequences on $\CC$ in $\MMM$.  

\begin{definition}
A {\it $\CC$-colored operad} in $\MMM$ 
is a symmetric sequence $\OO$ on $\CC$ in $\MMM$
with the following extra structure: 
\begin{enumerate}[i)]
\item For all $\big( \substack{t \\ \und{a}} \big) \in \CC^{n+1}$
and $\big( \substack{a_i \\ \und{b}_i} \big) \in \CC^{k_i+1}$
a composition 
\begin{subequations}
\begin{equation}
\gamma : \OO \big( \substack{t \\ \und{a}} \big) 
\otimes \OO \big( \substack{a_1 \\ \und{b}_1} \big) 
\otimes \dots
\otimes \OO \big( \substack{a_n \\ \und{b}_n} \big) 
\to \OO \big( \substack{t \\ \und{b}} \big) 
\end{equation}
where $\und{b} = (\und{b}_1, \dots \und{b}_n)$ is defined by concatenation. 
\item For all $t \in \CC$ a unit 
\begin{equation} \oone \in \OO \big( \substack{t \\ t} \big)~.
\end{equation}
\end{subequations}
\end{enumerate}
These structures are required to satisfy certain axioms
expressing associativity of $\gamma$, unitality of $\oone$
and compatibility between $\gamma$ and the symmetric action; 
for these and more details we refer to \cite{yau2016colored}. 

A morphism of operads $\phi : \OO \to \PP$ is a family of $\MMM$-morphisms
$\phi: \OO \big( \substack{t \\ \und{c}} \big) \to \PP \big( \substack{t \\ \und{c}} \big) $
that is equivariant with respect to the symmetric action, 
\begin{equation}
\PP_\sigma (\phi (o )) = \phi ( \OO_\sigma (o))
\end{equation}
for $o \in \OO \ipt{t}{\und{c}}$ and $\sigma \in \Sigma_n$, 
commutes with the composition, 
\begin{equation}
\gamma_\PP \big( \phi (o); \phi (o_1), \dots \phi (o_n)  \big) = \phi \big( \gamma_\OO (o; o_1, \dots o_n) \big)
\end{equation}
for $o \in \OO \big( \substack{t \\ \und{a}} \big) $ and $o_i \in \OO \big( \substack{a_i \\ \und{b}_i} \big) $, 
and is compatible with the unit, 
\begin{equation}
\oone_\PP = \phi (\oone_\OO) \in \PP \ipt{t}{t}~.
\end{equation}
The category of $\CC$-colored operads in $\MMM$ is denoted by $\Op_\CC(\MMM)$. 

More generally we define $\Op(\MMM)$ as the category of operads in $\MMM$, 
which are pairs $(\CC, \OO)$ where $\CC$ is a set of colors and $\OO$ is a $\CC$-colored operad in $\MMM$. 
A morphism $(f, \phi): (\CC , \OO) \to ( \DD , \PP )$ in $\Op (\MMM)$ 
is a map of colors $f: \CC \to \DD$ 
together with a morphism of $\CC$-colored operads $\phi : \OO \to f^*\PP$. 
Here, $f^* \PP$ is the pullback operad with 
$f^* \PP \ipt{t}{\und{c}} = \PP \ipt{f(t)}{f(\und{c})}$. 
\end{definition}

At this point, an example to illustrate this rather abstract definition is in order. 
First, a specialization of the previous definition: 
we call an operad $\OO$ colored over the singleton set $\ops$ an {\it uncolored operad}; 
we write $\OO(n) = \OO \big( \substack{ * \\ *^n } \big)$
where $*^n = (*, \dots *)$ is the sequence containing $*$ $n$ times. 

We can visualize elements of an operad with (directed and rooted) trees. 
An element $o \in \OO (n)$ will be a tree with $n$ inputs and one output, 
\begin{equation}
\begin{tikzpicture}
\draw (0,0.4) -- (0,0)  node [right=1pt] {\small{$o$}};
\draw (0,0) -- (-0.45,-0.3);
\draw (0,0) -- (-0.25,-0.3);
\draw (-0.15,-0.3) -- (0.35,-0.3) [dotted];
\draw (0,0) -- (0.45,-0.3);
\end{tikzpicture} 
\end{equation}
By grafting the trees we get a composition and permutation of the leaves gives a symmetric action. 
Let us illustrate this by constructing the (uncolored) {\it associative operad} in $\Set$ using tree diagrams. 

We start with a single tree with two inputs, 
\begin{equation}
\begin{tikzpicture}
\draw (0,0.25) -- (0,0);
\draw (0,0) -- (-0.2,-0.2);
\draw (0,0) -- (0.2,-0.2);
\end{tikzpicture} 
\end{equation}
which will represent multiplication in an algebra. 
Composition is done by grafting trees. 
For example, 
\begin{equation}
\gamma \big(
\begin{tikzpicture}
\draw (0,0.25) -- (0,0);
\draw (0,0) -- (-0.2,-0.2);
\draw (0,0) -- (0.2,-0.2);
\end{tikzpicture} 
~ ; ~
\begin{tikzpicture}
\draw (0,0.25) -- (0,0);
\draw (0,0) -- (-0.2,-0.2);
\draw (0,0) -- (0.2,-0.2);
\end{tikzpicture} 
~ , ~
\begin{tikzpicture}
\draw (0,0.25) -- (0,-0.1);
\end{tikzpicture} 
~
\big)
= 
\begin{tikzpicture}[baseline=-6]
\draw (0,0.25) -- (0,0);
\draw (0,0) -- (-0.2,-0.2);
\draw (0,0) -- (0.2,-0.2)  ;
\draw (-0.2,-0.2) -- (-0.4,-0.4) ;
\draw (-0.2,-0.2) -- (0,-0.4) ;
\end{tikzpicture}
\end{equation}
and
\begin{equation}
\gamma \big(
\begin{tikzpicture}
\draw (0,0.25) -- (0,0);
\draw (0,0) -- (-0.2,-0.2);
\draw (0,0) -- (0.2,-0.2);
\end{tikzpicture} 
~ ; ~
\begin{tikzpicture}
\draw (0,0.25) -- (0,-0.1);
\end{tikzpicture} 
~ , ~
\begin{tikzpicture}
\draw (0,0.25) -- (0,0);
\draw (0,0) -- (-0.2,-0.2);
\draw (0,0) -- (0.2,-0.2);
\end{tikzpicture}
~
\big)
= 
\begin{tikzpicture}[baseline=-6]
\draw (0,0.25) -- (0,0);
\draw (0,0) -- (-0.2,-0.2);
\draw (0,0) -- (0.2,-0.2);
\draw (0.2,-0.2) -- (0,-0.4);
\draw (0.2,-0.2) -- (0.4,-0.4);
\end{tikzpicture} .
\end{equation}
Here, we've already used the operadic unit $\oone$, which is the tree with one input
\begin{equation}
 \oone = ~ \begin{tikzpicture}
\draw (0,0.25) -- (0,-0.1);
\end{tikzpicture} ~.
\end{equation}

The symmetric action will permute the inputs, 
which we will label to keep track of this. 
So we write 
\begin{tikzpicture}[baseline=-6]
\draw (0,0.25) -- (0,0);
\draw (0,0) -- (-0.2,-0.2) node [below=-1pt] {\tiny{1}};
\draw (0,0) -- (0.2,-0.2) node [below=-1pt] {\tiny{2}};
\end{tikzpicture} 
for our generator 
\begin{tikzpicture}[baseline=-6]
\draw (0,0.25) -- (0,0);
\draw (0,0) -- (-0.2,-0.2);
\draw (0,0) -- (0.2,-0.2);
\end{tikzpicture}
and we have
\begin{equation}
\OO_{(12)} \big(
\begin{tikzpicture}[baseline=-6]
\draw (0,0.25) -- (0,0);
\draw (0,0) -- (-0.2,-0.2) node [below=-1pt] {\tiny{1}};
\draw (0,0) -- (0.2,-0.2) node [below=-1pt] {\tiny{2}};
\end{tikzpicture}
\big)
=
\begin{tikzpicture}[baseline=-6]
\draw (0,0.25) -- (0,0);
\draw (0,0) -- (-0.2,-0.2) node [below=-1pt] {\tiny{2}};
\draw (0,0) -- (0.2,-0.2) node [below=-1pt] {\tiny{1}};
\end{tikzpicture}
\end{equation}
representing the opposite multiplication. 
In general,
\begin{equation}
\OO_\sigma \big(
\begin{tikzpicture}[baseline=-6]
\draw (0,0.4) -- (0,0);
\draw (0,0) -- (-0.45,-0.3) node [below=-1pt] {\tiny{1}};
\draw (0,0) -- (-0.25,-0.3);
\draw (-0.15,-0.3) -- (0.35,-0.3) [dotted];
\draw (0,0) -- (0.45,-0.3) node [below=0.5pt] {\tiny{n}};
\draw (0,0) circle (4pt) [fill=white]; 
\end{tikzpicture} 
\big)
 = 
\begin{tikzpicture}[baseline=-6]
\draw (0,0.4) -- (0,0) ;
\draw (0,0) -- (-0.45,-0.3) node [below=-1pt] {\tiny{$\sigma^{-1}(1)$}};
\draw (0,0) -- (-0.25,-0.3);
\draw (-0.15,-0.3) -- (0.35,-0.3) [dotted];
\draw (0,0) -- (0.45,-0.3) node [below=-1pt] {\tiny{$\sigma^{-1}(n)$}};
\draw (0,0) circle (4pt) [fill=white]; 
\end{tikzpicture} 
\end{equation}
where $\begin{tikzpicture}[baseline=-6]
\draw (0,0.4) -- (0,0);
\draw (0,0) -- (-0.45,-0.3);
\draw (0,0) -- (-0.25,-0.3);
\draw (-0.15,-0.3) -- (0.35,-0.3) [dotted];
\draw (0,0) -- (0.45,-0.3) ;
\draw (0,0) circle (4pt) [fill=white]; 
\end{tikzpicture} $ represents any tree we can make by grafting. 

Compatibility of the composition and the symmetric action means that for example
\begin{equation}
\gamma \big(
\begin{tikzpicture}[baseline=-6]
\draw (0,0.25) -- (0,0);
\draw (0,0) -- (-0.2,-0.2) node [below=-1pt] {\tiny{2}};
\draw (0,0) -- (0.2,-0.2) node [below=-1pt] {\tiny{1}};
\end{tikzpicture} 
~ ; ~
\begin{tikzpicture}
\draw (0,0.25) -- (0,-0.1);
\end{tikzpicture} 
~ , ~
\begin{tikzpicture}[baseline=-6]
\draw (0,0.25) -- (0,0);
\draw (0,0) -- (-0.2,-0.2) node [below=-1pt] { \tiny{1}};
\draw (0,0) -- (0.2,-0.2) node [below=-1pt] {\tiny{2}};
\end{tikzpicture} 
~
\big)
= 
\begin{tikzpicture}[baseline=-6]
\draw (0,0.25) -- (0,0);
\draw (0,0) -- (-0.2,-0.2);
\draw (0,0) -- (0.2,-0.2)  node [below=-1pt] { \tiny{1}} ;
\draw (-0.2,-0.2) -- (-0.4,-0.4) node [below=-1pt] { \tiny{2}} ;
\draw (-0.2,-0.2) -- (0,-0.4) node [below=-1pt] { \tiny{3}} ;
\end{tikzpicture}
\end{equation}
i.e. the permutation $(12)$ in $\begin{tikzpicture}[baseline=-6]
\draw (0,0.25) -- (0,0);
\draw (0,0) -- (-0.2,-0.2) node [below=-1pt] {\tiny{2}};
\draw (0,0) -- (0.2,-0.2) node [below=-1pt] {\tiny{1}};
\end{tikzpicture} = \OO_{(12)} \big(
\begin{tikzpicture}[baseline=-6]
\draw (0,0.25) -- (0,0);
\draw (0,0) -- (-0.2,-0.2) node [below=-1pt] {\tiny{1}};
\draw (0,0) -- (0.2,-0.2) node [below=-1pt] {\tiny{2}};
\end{tikzpicture}
\big)$ descends to the inputs
and the operation \begin{tikzpicture}[baseline=-6]
\draw (0,0.25) -- (0,0);
\draw (0,0) -- (-0.2,-0.2) node [below=-1pt] { \tiny{1}};
\draw (0,0) -- (0.2,-0.2) node [below=-1pt] {\tiny{2}};
\end{tikzpicture} in the second slot ends up on the left. 

A general element we can now construct is a flat binary tree 
(a tree with two inputs and one output at every node) 
with its $n$ inputs labelled by $\sigma(1)$ to $\sigma(n)$
for any $\sigma \in \Sigma_n$. 
For example, we have the element
\begin{equation}
\begin{tikzpicture}
\draw (0,0.4) -- (0,0) node [right=1pt] {\small{}};
\draw (0,0) -- (-0.3,-0.3);
\draw (0,0) -- (0.3,-0.3);
\draw (-0.3,-0.3) -- (-0.50,-0.6) node [below=-2pt] { \tiny{2}};
\draw (-0.3,-0.3) -- (-0.10,-0.6) node [below=-2pt] { \tiny{5}};
\draw (0.3,-0.3) -- (0.10,-0.6);
\draw (0.3,-0.3) -- (0.50,-0.6) node [below=-2pt] { \tiny{1}};
\draw (0.10,-0.6) -- (-0.10,-0.9) node [below=-2pt] { \tiny{4}};
\draw (0.10,-0.6) -- (0.30,-0.9) node [below=-2pt] { \tiny{3}};
\end{tikzpicture} 
\end{equation}
in arity 5. 

We then implement the associativity relation
\begin{equation}
\begin{tikzpicture}[baseline=-6]
\draw (0,0.25) -- (0,0);
\draw (0,0) -- (-0.2,-0.2);
\draw (0,0) -- (0.2,-0.2)  ;
\draw (-0.2,-0.2) -- (-0.4,-0.4) ;
\draw (-0.2,-0.2) -- (0,-0.4) ;
\end{tikzpicture}
=
\begin{tikzpicture}[baseline=-6]
\draw (0,0.25) -- (0,0);
\draw (0,0) -- (-0.2,-0.2);
\draw (0,0) -- (0.2,-0.2);
\draw (0.2,-0.2) -- (0,-0.4);
\draw (0.2,-0.2) -- (0.4,-0.4);
\end{tikzpicture}
\end{equation}
for all possible input data. 
Using this relation, we can bring every tree into a standard form, say
\begin{equation}
\begin{tikzpicture}
\draw (0,0.25) -- (0,0);
\draw (0,0) -- (-0.2,-0.2) node[below left = -3pt] {\tiny{$\sigma(1)$}};
\draw (0,0) -- (0.2,-0.2);
\draw (0.2,-0.2) -- (0.0,-0.4); 
\draw (0.2,-0.2) -- (0.4,-0.4);
\draw (0.4,-0.4) -- (0.6,-0.6) [dotted];
\draw (0.7,-0.7) -- (0.5,-0.9); 
\draw (0.65,-0.65) -- (0.9,-0.9) node [below right = -3 pt] {\tiny{$\sigma(n)$}};
\end{tikzpicture}~. 
\end{equation}

We introduce another generator
\begin{tikzpicture}
\draw (0,0.25) -- (0,-0.1);
\draw (0,-0.1) circle (1.5pt) [fill=white]; 
\end{tikzpicture}, 
which has zero inputs
and represents the unit in the algebra. 
Imposing the relations 
\begin{equation}
\begin{tikzpicture}[baseline=-6]
\draw (0,0.25) -- (0,0);
\draw (0,0) -- (-0.2,-0.2);
\draw (0,0) -- (0.2,-0.2);
\draw (-0.2,-0.2) circle (1.5pt) [fill=white]; 
\end{tikzpicture} 
=
\begin{tikzpicture}
\draw (0,0.25) -- (0,-0.1);
\end{tikzpicture}
=
\begin{tikzpicture}[baseline=-6]
\draw (0,0.25) -- (0,0);
\draw (0,0) -- (-0.2,-0.2);
\draw (0,0) -- (0.2,-0.2);
\draw (0.2,-0.2) circle (1.5pt) [fill=white]; 
\end{tikzpicture} 
\end{equation}
expresses this unitality. 

We now define $\As$ to be the operad generated by 
\begin{tikzpicture}
\draw (0,0.25) -- (0,0);
\draw (0,0) -- (-0.2,-0.2);
\draw (0,0) -- (0.2,-0.2);
\end{tikzpicture} 
and 
\begin{tikzpicture}
\draw (0,0.25) -- (0,-0.1);
\draw (0,-0.1) circle (1.5pt) [fill=white]; 
\end{tikzpicture}
with the associativity and unitality relations imposed. 
With our standard tree form, 
we find that $\As (n)$ contains $n!$ elements. 

In the construction of $\As$ we used the general result 
of being able to present an operad by generators and relations. 
More precisely, given a sequence $X$ 
one can construct the {\it free operad} $F(X)$. 
The relation then gives us two points in $F(X)$, 
and we can define the quotient, 
or in categorical language the coequalizer of the corresponding diagram in $\Op_\CC(\MMM)$. 
We again refer to \cite{yau2016colored} for details on these constructions. 

So we have constructed the associative operad as
\begin{equation}
\xymatrix@C=2.5em
{ 
F \big(
\{ u_1, u_2 \} [1] \cup \{ a \} [3]
\big)
\ar@<0.5ex>[r]^-{r_1}  
\ar@<-0.5ex>[r]_-{r_2}
~&~
F \big(
\begin{tikzpicture}
\draw (0,0.25) -- (0,0);
\draw (0,0) -- (-0.2,-0.2);
\draw (0,0) -- (0.2,-0.2);
\end{tikzpicture} , 
\begin{tikzpicture}
\draw (0,0.25) -- (0,-0.1);
\draw (0,-0.1) circle (1.5pt) [fill=white]; 
\end{tikzpicture} \,
\big) 
\ar@{-->}[r]  
~&~  
\As 
}
\end{equation}
in $\Op_\ops (\Set)$. 
Here, 
\begin{equation}
 r_1 ( a) = \begin{tikzpicture}[baseline=-6]
\draw (0,0.25) -- (0,0);
\draw (0,0) -- (-0.2,-0.2);
\draw (0,0) -- (0.2,-0.2);
\draw (-0.2,-0.2) -- (-0.4,-0.4);
\draw (-0.2,-0.2) -- (0,-0.4);
\end{tikzpicture} ~ , \quad 
r_2 (a) = \begin{tikzpicture}[baseline=-6]
\draw (0,0.25) -- (0,0);
\draw (0,0) -- (0.2,-0.2);
\draw (0,0) -- (-0.2,-0.2);
\draw (0.2,-0.2) -- (0,-0.4);
\draw (0.2,-0.2) -- (0.4,-0.4);
\end{tikzpicture}
\end{equation}
which imposes associativity, 
and 
\begin{equation}
 r_1 (u_1) = \begin{tikzpicture}[baseline=-6]
\draw (0,0.25) -- (0,0);
\draw (0,0) -- (-0.2,-0.2);
\draw (0,0) -- (0.2,-0.2);
\draw (-0.2,-0.2) circle (1.5pt) [fill=white]; 
\end{tikzpicture} ~ , \quad
r_1 (u_2) = \begin{tikzpicture}[baseline=-6]
\draw (0,0.25) -- (0,0);
\draw (0,0) -- (-0.2,-0.2);
\draw (0,0) -- (0.2,-0.2);
\draw (0.2,-0.2) circle (1.5pt) [fill=white]; 
\end{tikzpicture}  ~ , \quad 
r_2 (u_i) = \begin{tikzpicture}
\draw (0,0.25) -- (0,-0.1);
\end{tikzpicture}
\end{equation}
implementing unitality. 
The $[3]$ in $\{ a \} [3]$ 
indicates that we are mapping to operations in arity 3, 
i.e. $\begin{tikzpicture}[baseline=-6]
\draw (0,0.25) -- (0,0);
\draw (0,0) -- (-0.2,-0.2);
\draw (0,0) -- (0.2,-0.2);
\draw (0.2,-0.2) -- (0,-0.4);
\draw (0.2,-0.2) -- (0.4,-0.4);
\end{tikzpicture} \in F \big(
\begin{tikzpicture}
\draw (0,0.25) -- (0,0);
\draw (0,0) -- (-0.2,-0.2);
\draw (0,0) -- (0.2,-0.2);
\end{tikzpicture} 
\big) (3)$. 

We can construct operads expressing other algebraic structures in a similar way. 
For example, the {\it Lie operad} $\Lie$ is the uncolored operad generated by 
\begin{tikzpicture}
\draw (0,0.25) -- (0,0);
\draw (0,0) -- (-0.2,-0.2);
\draw (0,0) -- (0.2,-0.2);
\end{tikzpicture} 
which now represents the Lie bracket, 
with anticommutativity 
$\begin{tikzpicture}[baseline=-6]
\draw (0,0.25) -- (0,0);
\draw (0,0) -- (-0.2,-0.2) node [below=-1pt] {\tiny{2}};
\draw (0,0) -- (0.2,-0.2) node [below=-1pt] {\tiny{1}};
\end{tikzpicture} = - \begin{tikzpicture}[baseline=-6]
\draw (0,0.25) -- (0,0);
\draw (0,0) -- (-0.2,-0.2) node [below=-1pt] {\tiny{1}};
\draw (0,0) -- (0.2,-0.2) node [below=-1pt] {\tiny{2}};
\end{tikzpicture}$
and the Jacobi relation 
$\begin{tikzpicture}[baseline=-6]
\draw (0,0.25) -- (0,0);
\draw (0,0) -- (0.2,-0.2);
\draw (0,0) -- (-0.2,-0.2) node [below left=-1pt] { \tiny{1}} ;
\draw (0.2,-0.2) -- (0,-0.4) node [below =-1pt] { \tiny{2}} ;
\draw (0.2,-0.2) -- (0.4,-0.4) node [below =-1pt] { \tiny{3}} ;
\end{tikzpicture}
+
\begin{tikzpicture}[baseline=-6]
\draw (0,0.25) -- (0,0);
\draw (0,0) -- (0.2,-0.2);
\draw (0,0) -- (-0.2,-0.2) node [below left=-1pt] { \tiny{2}} ;
\draw (0.2,-0.2) -- (0,-0.4) node [below =-1pt] { \tiny{3}} ;
\draw (0.2,-0.2) -- (0.4,-0.4) node [below =-1pt] { \tiny{1}} ;
\end{tikzpicture}
+
\begin{tikzpicture}[baseline=-6]
\draw (0,0.25) -- (0,0);
\draw (0,0) -- (0.2,-0.2);
\draw (0,0) -- (-0.2,-0.2) node [below left=-1pt] { \tiny{3}} ;
\draw (0.2,-0.2) -- (0,-0.4) node [below =-1pt] { \tiny{1}} ;
\draw (0.2,-0.2) -- (0.4,-0.4) node [below =-1pt] { \tiny{2}} ;
\end{tikzpicture}
= 0
$. 
Note that since we need addition and subtraction, 
this construction doesn't make sense in $\Set$, so we instead work in $\Vec_k$; 
the trees now form a basis for our vector space, 
and we extend our operations linearly. 
A variation on the Lie operad is the {\it unital Lie operad} $\uLie$, 
which has an extra generator 
\begin{tikzpicture}
\draw (0,0.25) -- (0,-0.1);
\draw (0,-0.1) circle (1.5pt) [fill=white]; 
\end{tikzpicture}
with the relation 
$\begin{tikzpicture}[baseline=-6]
\draw (0,0.25) -- (0,0);
\draw (0,0) -- (-0.2,-0.2);
\draw (0,0) -- (0.2,-0.2);
\draw (-0.2,-0.2) circle (1.5pt) [fill=white]; 
\end{tikzpicture} = 0$. 

Likewise, one can present the (uncolored) {\it Poisson operad} $\Pois$ in $\Vec_k$
by starting with a Lie generator 
\begin{tikzpicture}
\draw (0,0.25) -- (0,0);
\draw (0,0) -- (-0.2,-0.2);
\draw (0,0) -- (0.2,-0.2);
\end{tikzpicture} 
and a generator
\begin{tikzpicture}
\draw (0,0.25) -- (0,0);
\draw (0,0) -- (-0.2,-0.2);
\draw (0,0) -- (0.2,-0.2);
\draw (0,0) circle (1.5pt) [fill=white]; 
\end{tikzpicture} 
representing the commutative multiplication, 
and imposing anticommutativity and the Jacobi identity for the Lie generator, 
commutativity  
$\begin{tikzpicture}[baseline=-6]
\draw (0,0.25) -- (0,0);
\draw (0,0) -- (-0.2,-0.2) node [below=-1pt] {\tiny{2}};
\draw (0,0) -- (0.2,-0.2) node [below=-1pt] {\tiny{1}};
\draw (0,0) circle (1.5pt) [fill=white]; 
\end{tikzpicture} =  \begin{tikzpicture}[baseline=-6]
\draw (0,0.25) -- (0,0);
\draw (0,0) -- (-0.2,-0.2) node [below=-1pt] {\tiny{1}};
\draw (0,0) -- (0.2,-0.2) node [below=-1pt] {\tiny{2}};
\draw (0,0) circle (1.5pt) [fill=white]; 
\end{tikzpicture}$, 
and distributivity 
$ \begin{tikzpicture}[baseline=-6]
\draw (0,0.25) -- (0,0);
\draw (0,0) -- (-0.2,-0.2)node [below=-1pt] {\tiny{1}};
\draw (0,0) -- (0.2,-0.2);
\draw (0.2,-0.2) -- (0,-0.4)node [below=-1pt] {\tiny{2}};
\draw (0.2,-0.2) -- (0.4,-0.4)node [below=-1pt] {\tiny{3}};
\draw (0.2,-0.2) circle (1.5pt) [fill=white];
\end{tikzpicture} 
= 
\begin{tikzpicture}[baseline=-6]
\draw (0,0.25) -- (0,0);
\draw (0,0) -- (-0.2,-0.2);
\draw (0,0) -- (0.2,-0.2) node [below=-1pt] {\tiny{3}};
\draw (-0.2,-0.2) -- (-0.4,-0.4) node [below=-1pt] {\tiny{1}};
\draw (-0.2,-0.2) -- (0,-0.4) node [below=-1pt] {\tiny{2}};
\draw (0,0) circle (1.5pt) [fill=white];
\end{tikzpicture}
+ 
\begin{tikzpicture}[baseline=-6]
\draw (0,0.25) -- (0,0);
\draw (0,0) -- (-0.2,-0.2) node [below=-1pt] {\tiny{2}};
\draw (0,0) -- (0.2,-0.2);
\draw (0.2,-0.2) -- (0,-0.4) node [below=-1pt] {\tiny{1}};
\draw (0.2,-0.2) -- (0.4,-0.4) node [below=-1pt] {\tiny{3}};
\draw (0,0) circle (1.5pt) [fill=white];
\end{tikzpicture} 
$
. 

An example of a colored operad in $\Set$ 
is the {\it diagram operad} of a small category $\CCC$, $\Diag_\CCC$. 
Its colors are the objects in $\CCC$, $\CCC_0$. 
For $c,t \in \CCC_0$ we define 
\begin{equation}
\Diag_\CCC \big( \substack{t \\ c } \big) = \CCC (c, t)
\end{equation}
as the set of morphisms from $c$ to $t$ in $\CCC$, 
and $\Diag_\CCC \big( \substack{t \\ \und{c} } \big) = \emptyset$ if $|\und{c}| \neq 1$, 
so $\Diag_\CCC$ is only non-empty in arity 1. 
Since $\Sigma_1 = \{ e \}$, there is no (non-trivial) symmetric action. 
Composition 
\begin{equation}
\gamma: \Diag_\CCC \big( \substack{t \\ a } \big)
\times \Diag_\CCC \big( \substack{a \\ b } \big)
\to
\Diag_\CCC \big( \substack{t \\ b } \big)
\end{equation}
is the usual composition of morphisms in $\CCC$
and the unit is 
\begin{equation}
 \oone = \id_t \in \Diag_\CCC \ipt{t}{t}~.
 \end{equation}

We can expand this operad to include multiple maps with the same target: 
still working with a small category $\CCC$
we define the $\CCC_0$-colored operad $\MDiag_\CCC$ as 
\begin{equation}
\MDiag \ipt{t}{\und{c}} = \CCC ( \und{c}, t ) = \prod_{i=1}^n \CCC  (c_i, t)
\end{equation}
for $\ipt{t}{\und{c}} \in \CCC_0^{n+1}$. 
We write $\und{f} = (f_1, \dots, f_n) \in \MDiag \ipt{t}{\und{c}}$ 
with $f_i: c_i \to t$. 
The symmetric action is 
\begin{equation}
\und{f} \mapsto \und{f} \sigma = (f_{\sigma(1)}, \dots, f_{\sigma(n)} ) 
\end{equation}
for $\und{f} \in \MDiag_\CCC \ipt{t}{\und{c}}$ and $\sigma \in \Sigma_n$. 
Composition is induced by composition of maps in $\CCC$, 
\begin{equation} 
(\und{f} ; \und{g}_1, \dots, \und{g}_n) 
\mapsto   (f_1 g_{11}, \dots, f_i g_{ij}, \dots, f_n g_{nk_n} )
\end{equation}
for $\und{f} \in \MDiag \ipt{t}{\und{a}}$ and $\und{g}_i \in \MDiag \ipt{a_i}{\und{b}_i}$. 
We write 
\begin{equation}\und{f} ( \und{g}_1, \dots, \und{g}_n )
=  (f_1 g_{11}, \dots, f_n g_{nk_n} )~.
\end{equation}
The unit is
\begin{equation}
 \oone = \id_t \in \MDiag_\CCC \ipt{t}{t}~.
 \end{equation}

\subsection{Algebras over operads}\label{algebras}

We study operads because we are interested in their algebras; 
the not-defined notion of ``algebraic category'' mentioned in Section \ref{AQFT}
will be the category of algebras over an operad. 

For a set $\CC$, a {\it $\CC$-colored object} $X$ in $\MMM$
is an assignment $c \mapsto X_c \in \MMM$ for all $c \in \CC$. 
The category of $\CC$-colored objects 
is isomorphic to the functor category $\MMM^\CC$
if we again view $\CC$ as a discrete category. 

\begin{definition}
An {\it algebra} over an operad $\OO \in \Op_\CC(\MMM)$
is a $\CC$-colored object $A$ with an action of the operad
\begin{equation}
 \alpha : \OO \ipt{t}{\und{c}} \otimes A_{c_1} \otimes \dots \otimes A_{c_n} \to A_t~.
 \end{equation}
This $\alpha$ is required to satisfy compatibility axioms
with respect to the composition, unit and symmetric action on $\OO$; 
we again refer to \cite{yau2016colored} for the details. 

A {\it morphism} of $\OO$-algebras $\kappa: A \to B$ 
is a family of maps $A_c \to B_c$
that is compatible with respect to the $\OO$-action, 
\begin{equation} 
\alpha_B \big( o ; \kappa(a_1) , \dots, \kappa (a_n) \big)
= \kappa \big( \alpha_A (o ; a_1, \dots, a_n) \big)
\end{equation}
for $o \in \OO \ipt{t}{\und{c}}$ and $a_i \in A_{c_i}$. 
We denote the category of algebras over $\OO$ by $\Alg ( \OO )$. 
\end{definition}

We can now elucidate the name of the associative operad. 
For concreteness, choose $\Vec_k$ as our monoidal category; 
the construction in Section \ref{operads} immediately generalizes to $\Vec_k$
by defining $\As(n)_{\Vec_k}$ to be the vector space
with the trees in $\As(n)_\Set$ as a basis
and extending all operations linearly. 

Since $\As$ is uncolored, an algebra over $\As$
is an object $A = A_* \in M$ 
with an $\As$-action 
\begin{equation}
\alpha : \As(n) \otimes A^{\otimes n} \to A~.
\end{equation}
Since $\As$ is generated by \begin{tikzpicture}
\draw (0,0.25) -- (0,0);
\draw (0,0) -- (-0.2,-0.2);
\draw (0,0) -- (0.2,-0.2);
\end{tikzpicture} and 
\begin{tikzpicture}
\draw (0,0.25) -- (0,-0.1);
\draw (0,-0.1) circle (1.5pt) [fill=white]; 
\end{tikzpicture}, 
this action is determined by multiplication $\mu = 
\alpha (\begin{tikzpicture}
\draw (0,0.25) -- (0,0);
\draw (0,0) -- (-0.2,-0.2);
\draw (0,0) -- (0.2,-0.2);
\end{tikzpicture} ; - , - )$: 
\begin{equation}
\begin{aligned} 
\mu : A \otimes A  &\to  A~,  \\
 a_1 \otimes a_2  &\mapsto  a_1 \cdot a_2 &
\end{aligned}
\end{equation}
and a unit $e = \alpha(\begin{tikzpicture}
\draw (0,0.25) -- (0,-0.1);
\draw (0,-0.1) circle (1.5pt) [fill=white]; 
\end{tikzpicture})$: 
\begin{equation}
\begin{aligned}
e: A^{\otimes 0} = k & \to A  \\
1 & \mapsto e~.
\end{aligned}
\end{equation}
The associativity relation implies that
\begin{equation} (a_1 \cdot a_2) \cdot a_3 = a_1 \cdot (a_2 \cdot a_3) \end{equation}
so our multiplication is associative, 
and the unitality axioms imply that 
\begin{equation} a \cdot e = a = e \cdot a\end{equation}
so $e$ is a unit. 
We see that algebras over $\As$ are exactly associative algebras with unit. 

Similarly, algebras over $\Lie$ are Lie algebras, 
with $[-,-] = \alpha (\begin{tikzpicture}
\draw (0,0.25) -- (0,0);
\draw (0,0) -- (-0.2,-0.2);
\draw (0,0) -- (0.2,-0.2);
\end{tikzpicture} ; - , - )$ the Lie bracket, 
and algebras over $\Pois$ are Poisson algebras, 
with $\{ -,- \} = \alpha (\begin{tikzpicture}
\draw (0,0.25) -- (0,0);
\draw (0,0) -- (-0.2,-0.2);
\draw (0,0) -- (0.2,-0.2);
\end{tikzpicture} ; - , - )$ the Poisson bracket
and $ \mu = \alpha (\begin{tikzpicture}
\draw (0,0.25) -- (0,0);
\draw (0,0) -- (-0.2,-0.2);
\draw (0,0) -- (0.2,-0.2);
\draw (0,0) circle (1.5pt) [fill=white]; 
\end{tikzpicture} ; - , - )$ the (commutative) multiplication. 

An algebra over $\Diag_\CCC$ is an assignment
\begin{equation}A: \CCC_0 \to \MMM\end{equation}
with a $\Diag_\CCC$-action 
\begin{equation}\alpha: \Diag_\CCC(c,t) \otimes A_c \to A_t  \end{equation}
or in other words an assignment $\CCC (c , t) \to \MMM (A_c , A_t)$. 
Compatibility exactly means that these assignments respect composition and units, 
so the algebras over $\Diag_\CCC$ are the functors from $\CCC$ to $\MMM$, 
$\Alg ( \Diag_\CCC) \cong \MMM^\CCC$. 

Given a morphism of operads $(f, \phi): (\CC, \OO) \to (\DD, \PP)$
one can pull back algebras: 
for $(A,\alpha) \in \Alg (\PP)$, 
\begin{equation}(f, \phi)^* (A,\alpha) = (f^*A, \phi^* \alpha )\end{equation}
where
\begin{equation} (f^*A)_c = A_{f(c)}\end{equation}
and 
\begin{equation} ( \phi^* \alpha ) ( o ; a_1, \dots a_n) = \alpha (\phi (o) ; a_1, \dots a_n) \end{equation}
is the $\OO$-action on $f^*A$
where $o \in  \OO \ipt{t}{\und{c}}$ and $a_i \in (f^*A)_{c_i} = A_{f(c_i)}$. 

An important result is that taking the pullback is a functor, 
and that this functor has a left adjoint which we denote by $ (f,\phi)_! $: 
\begin{equation}\xymatrix{
\Alg(\OO)
\ar@<0.7ex>[r]^{(f,\phi)_! }
& 
\Alg(\PP)
\ar@<0.7ex>[l]^{(f,\phi)^*}
}~. \end{equation}
We will use this extensively in Section \ref{adjunctions}.

\section{Operads in field theory}\label{operadsinFT}

\subsection{An operadic definition of field theory}\label{operadicFT}

We are now in a position to fix half of Definition \ref{defFT1}: 
we can replace ``algebraic category'' by ``category of algebras over an operad''. 
For the second part, commutativity, we take the following general approach. 

For an uncolored operad $\PP$, we pick out two operations of arity 2, $r_1, r_2 \in \PP (2)$. 
We then say that two elements $x,y \in (A, \alpha) \in \Alg (\PP)$ {\it commute}
if $\alpha ( r_1 ; x,y) = \alpha (r_2 ; x,y)$. 
We call $(\PP, r_i)$ a {\it bipointed operad} and we write $\Op^{2pt} (\MMM)$ 
for the category of (uncolored) bipointed operads in $\MMM$. 
A morphism of bipointed operads $\phi: (\PP, r_i) \to (\QQ, s_i)$ 
is a morphism of operads $\phi: \PP \to \QQ$
that preserves the chosen points: $\phi (r_i) = s_i$. 

As an example, we can choose $r_1 = \begin{tikzpicture}[baseline=-6]
\draw (0,0.25) -- (0,0);
\draw (0,0) -- (-0.2,-0.2) node [below=-1pt] {\tiny{1}};
\draw (0,0) -- (0.2,-0.2) node [below=-1pt] {\tiny{2}};
\end{tikzpicture}$ and $r_2 = \begin{tikzpicture}[baseline=-6]
\draw (0,0.25) -- (0,0);
\draw (0,0) -- (-0.2,-0.2) node [below=-1pt] {\tiny{2}};
\draw (0,0) -- (0.2,-0.2) node [below=-1pt] {\tiny{1}};
\end{tikzpicture}$ in $\As$, expressing regular commutativity $x \cdot y = y \cdot x$. 
Likewise, we can consider $[x,y] = 0$ in Lie algebras, 
and $\{ x , y \} = 0$ in Poisson algebras. 

We can now properly define our notion of field theory. 

\begin{definition}
Given an uncolored bipointed operad $(\PP, r_i)$, 
a {\it field theory} of type $(\PP, r_i)$ on an orthogonal category $\ovr{\CCC}$
is a functor 
\begin{subequations}
\begin{equation} \AA : \CCC \to \Alg (\PP) \end{equation}
such that if $(f_1: c_1 \to t) \perp (f_2: c_2 \to t)$, 
$\AA(f_1) \big( \AA(c_1) \big)$ and $\AA(f_2) \big( \AA(c_2) \big)$ commute in $\AA(t)$: 
\begin{equation} \alpha_t \big( r_1 ; \AA(f_1) ( x ) , \AA(f_2) ( y ) \big) = \alpha_t \big( r_2 ;\AA(f_1) ( x ) , \AA(f_2) ( y ) \big)  \end{equation}
for any $x \in \AA (c_1)$, $y \in \AA (c_2) $. 
We write $\FT ( \ovr{\CCC}, \PP, r_i)$ for the category
of field theories of type $(\PP, r_i)$ on $\ovr{\CCC}$. 
\end{subequations}
\end{definition}

As mentioned in Section \ref{AQFT} 
the time-slice property can be implemented by localization of the category $\CCC$. 
So we just focus on Einstein causality in the definition. 

\begin{example}
A {\it quantum field theory} is a field theory of type $(\As, \mu - \mu^{\text{op}} = 0)$
where $\mu^{\text{op}}$ is the opposite multiplication. 
So its algebras of observables are associative algebras, 
and algebras associated to spacelike separated subspace-times commute in the usual sense. 
We write $\QFT (\ovr{\CCC}) = \FT ( \ovr{\CCC}, \As, \mu - \mu^{\text{op}} = 0)$
for the category of quantum field theories. 

A {\it classical field theory} is a field theory of type $(\Pois, \{ , \} = 0)$, 
and a {\it linear field theory} is a field theory of type $( \uLie , [,] = 0)$. 
So we write $\CFT (\ovr{\CCC}) = \FT ( \ovr{\CCC}, \Pois, \{ , \} = 0)$
and $\LFT (\ovr{\CCC}) = \FT ( \ovr{\CCC}, \uLie , [,] = 0)$. 
\end{example}

\subsection{Field theory operads}\label{FToperads}

We have corrected the vagueness in our initial definition of field theories, 
but our definition as functors satisfying a certain property is still not very natural: 
we would like to have Einstein causality built in into the structure of the operad. 
In this section, we do this by coloring our operad with the space-time category
and proving that field theories are exactly algebras over a quotient of this operad. 

\begin{definition}
The {\it coloring} of an (uncolored) operad $\PP$
with a small category $\CCC$ 
is the $\CCC_0$-colored operad $\PP_\CCC$, 
where 
\begin{equation} \PP_\CCC \ipt{t}{\und{c}} 
= \coprod_{\und{f} \in \CCC(\und{c},t)}
\PP \big( | \und{c} | \big)  \end{equation}
for $t, c_i \in \CCC_0$ 
(recall that $\CCC ( \und{c}, t ) = \prod_{i=1}^n \CCC  (c_i, t) $). 
Concretely it contains elements $(\und{f}, p)$ 
with $\und{f} \in \CCC(\und{c},t)$ and $p \in \PP \big( | \und{c} | \big)$. 
On elements, the symmetric action is 
\begin{equation} (\PP_\CCC )_\sigma \big( \und{f},p \big) = \big( \und{f} \sigma , \PP_\sigma p \big)~. \end{equation}
Composition is defined as 
\begin{flalign*}
\gamma_{\PP_\CCC} \big( (\und{f}, p) ; (\und{g}_1, p_1) , \dots, & (\und{g}_n , p_n) \big) \\
& = \big( \und{f} (\und{g}_1, \dots, \und{g}_n) , \gamma_\PP ( p ; p_1, \dots, p_n) 
\big)
\end{flalign*}
and the unit is 
\begin{equation} \oone_{\PP_\CCC} = (\id_t , \oone_\PP ) \in \PP_\CCC \ipt{t}{t} . \end{equation}   
\end{definition}

Concretely, in $\Vec_k$ we have 
\begin{equation} \PP_\CCC \ipt{t}{\und{c}} 
= \bigoplus_{\und{f} \in \CCC(\und{c},t) } \PP \big( | \und{c} | \big)  \end{equation}
and in $\Set$
\begin{equation} \PP_\CCC \ipt{t}{\und{c}} 
= \coprod_{\und{f} \in \CCC(\und{c},t)} \PP \big( | \und{c} | \big) 
= \CCC(\und{c},t) \times \PP \big( | \und{c} | \big)~. \end{equation}
Recalling that $\MDiag \ipt{t}{\und{c}} = \CCC(\und{c},t)$
we find that $\PP_\CCC = \MDiag \times \PP$ is the arity-wise product of operads in $\Set$. 

The $\CCC$-coloring of $\PP$ is a natural object for us to study at this point, 
as is evidenced by the following result. 

\begin{lemma}\label{algaux}
Let $\PP$ be an uncolored operad 
and $\CCC$ be a small category. 
Then, we have an isomorphism of categories
\begin{equation} \Alg (\PP_\CCC) \cong \big( \Alg (\PP) \big)^\CCC\end{equation}
between the algebras over $\PP_\CCC$ 
and the functors from $\CCC$ to $\Alg (\PP)$. 
\end{lemma}

For the proof we refer to \cite{Bruinsma:2018knq}. 
Here, we will note that an algebra $A$ over $\PP_\CCC$ 
assigns to a color $c \in \CCC_0$ an object $A_c$, 
which is how the corresponding functor $F_A: \CCC \to \Alg(\PP)$ acts on objects. 
These $A_c$ are naturally $\PP$-algebras 
by considering the $\PP_\CCC$-action of $(\id_c , p)$ for $p \in \PP$. 
Lastly, $F_A(f) = \gamma( (f, \oone_\PP) ; - )$ gives $F_A$ on morphisms, 
and $F_A$ is then a functor by the operad axioms. 

To encode Einstein causality into our operad, 
we need to take a quotient. 
We now consider a bipointed operad $(\PP, r_i)$
and an orthogonal category $\ovr{\CCC} = (\CCC, \perp) $. 
Write 
\begin{equation}
\begin{aligned}
 \perp \ipt{t}{c_1, c_2} & = \big\{ (f_1: c_1 \to t, f_2 : c_2 \to t) | f_1 \perp f_2 \big\}  \\
 & = \perp \cap ~ \CCC( c_1, c_2 ; t) 
\end{aligned}
\end{equation}
and define the sequence
\begin{equation} \RR_\perp \ipt{t}{\und{c}}= \coprod_{\und{f} \in \perp \ipt{t}{\und{c}} } \ops~.  \end{equation}
Note that $\perp \ipt{t}{\und{c}} = \emptyset$ if $|\und{c}| \neq 2$
so this sequence is concentrated in arity 2. 
We can now define two maps of sequences
\begin{equation} r_i :  \RR_\perp \ipt{t}{\und{c}} \to \PP_\CCC \ipt{t}{\und{c}} \end{equation}
sending $(\und{f}, *)$ to $(\und{f} , r_1)$ and $(\und{f} , r_2)$ respectively, 
and we define the coequalizer in $\Op_{\CCC_0}(\MMM)$ of the corresponding maps, 
\begin{equation}
\kern-7pt\xymatrix@C=2.5em
{ 
F \big(
\RR_\perp 
\big)
\ar@<0.5ex>[r]^-{r_1}  
\ar@<-0.5ex>[r]_-{r_2}
~&~
\PP_\CCC
\ar@{-->}[r]  
~&~  
\PP_{\ovr{\CCC}}^r
}
\end{equation}
where we recall that $F \big( \RR_\perp \big)$
is the free operad generated by the sequence $\RR_\perp$. 

Our quotient of the coloring of an operad $\PP_{\ovr{\CCC}}^r$
is exactly the right object to get theories satisfying Einstein causality: 

\begin{proposition}\label{algFT}
Let $(\PP, r_i)$ be an uncolored bipointed operad 
and $\ovr{\CCC}$ be an orthogonal category. 
Then we have an isomorphism of categories
\begin{equation} \Alg ( \PP_{\ovr{\CCC}}^r) \cong \FT ( \ovr{\CCC}, \PP, r_i) \end{equation}
between the algebras over $\PP_{\ovr{\CCC}}^r$
and the field theories of type $(\PP, r_i)$ on $\ovr{\CCC}$. 
\end{proposition}

We again omit the proof, referring to \cite{Bruinsma:2018knq}. 
Note that this is a refinement of the previous lemma \ref{algaux}, and the proof is similar: 
we have to additionally note that our taking the quotient exactly enforces Einstein causality. 

The assignment $(\ovr{\CCC} , (\PP, r_i)) \mapsto \PP_{\ovr{\CCC}}^r$ is functorial in both arguments: 
\begin{subequations}
\begin{enumerate}[i)]
\item a morphism of orthogonal categories $F: \ovr{\CCC} \to \ovr{\DDD}$
gives rise to a map 
\begin{equation}
\begin{aligned}
 \PP_F^r :  \PP_{\ovr{\CCC}}^r & \longrightarrow \PP_{\ovr{\DDD}}^r~,   \\
 [\und{f}, p] &  \longmapsto [F(\und{f}) , p]~; 
\end{aligned} 
\end{equation}
\item a morphism of uncolored bipointed operads $\phi: (\PP, r_i) \to (\QQ, s_i)$ 
gives rise to a map
\begin{equation}
\begin{aligned}
 \phi_{\ovr{\CCC}}  :  \PP_{\ovr{\CCC}}^r & \longrightarrow \QQ_{\ovr{\CCC}}^s~,  \\
  [\und{f}, p] &  \longmapsto [\und{f} , \phi (p)]~; 
\end{aligned}
\end{equation} 
\item with such $F$ and $\phi$ we get the composition
\begin{equation}\phi_F : ~
\PP_{\ovr{\CCC}}^r 
\longrightarrow \QQ_{\ovr{\DDD}}^s  \end{equation} 
where $\phi_F 
= \QQ_F^s \circ \phi_{\ovr{\CCC}}  
= \phi_{\ovr{\DDD}}   \circ \PP_F^r $. 
\end{enumerate}
This functoriality will be used in the next section. 
\end{subequations}

\section{Adjunctions}\label{adjunctions}

Recall the statement at the end of Section \ref{algebras}:
for an operad map 
\begin{equation} (f , \phi ) : (\CC , \PP) \to (\DD , \QQ) \end{equation}
we have an adjunction 
\begin{equation}\kern-7pt\xymatrix{
\Alg(\PP)
\ar@<0.7ex>[r]^{(f,\phi)_! }
& 
\Alg(\QQ)
\ar@<0.7ex>[l]^{(f,\phi)^*}
} \end{equation}
where $(f , \phi)^*$ is the pullback. 
Our construction of $\PP_{\ovr{\CCC}}^r$ is functorial, 
which allows for two special types of maps of operads 
as mentioned in Section \ref{FToperads}: 
those coming from a change of orthogonal category $F: \ovr{\CCC} \to \ovr{\DDD}$
and those coming from a change of uncolored bipointed operad $\phi : (\PP,r_i) \to (\QQ,s_i)$. 

We will now work with these constructions to make some general statements about field theories. 
For all proofs of statements in this section we refer to \cite{Bruinsma:2018knq}.

\subsection{Change of color adjunctions}\label{changecolor}

A functor of orthogonal categories 
\begin{equation}F: \ovr{\CCC} \to \ovr{\DDD}\end{equation}
defines a map of the sets of objects $\CCC_0 \to \DDD_0$ 
which we also denote by $F$. 
From $F$, we get a map of operads 
\begin{equation} (F , \PP_F^r )  : ( \CCC_0 , \PP_{\ovr{\CCC}}^r )  \to ( \DDD_0 , \PP_{\ovr{\DDD}}^r ) \end{equation}
for any bipointed operad $(\PP , r_i)$. 
In turn, we get an adjunction 
\begin{equation}\kern-7pt\xymatrix{
\FT ( \ovr{\CCC}, \PP, r_i)
\ar@<0.7ex>[r]^{ (F , \PP_F^r )_! }
& 
\FT ( \ovr{\DDD}, \PP, r_i)
\ar@<0.7ex>[l]^{ (F , \PP_F^r )^*}
}\!\!\!\!\!. \end{equation}

From the construction of the pull-back 
we get a more explicit expression for $(F , \PP_F^r )^*$. 
Recalling that 
\begin{equation}\FT ( \ovr{\DDD}, \PP, r_i) \subseteq 
\big( \Alg (\PP) \big)^\DDD\end{equation}
we find that  
\begin{equation}(F ,  \PP_F^r )^* = F^*: \FT ( \ovr{\DDD}, \PP, r_i) \to \FT ( \ovr{\CCC}, \PP, r_i)\end{equation} 
is the regular pull-back, 
i.e.
\begin{equation}(F^*A)_c = A_{F(c)}\end{equation}
and 
\begin{equation} (F^* \alpha) ([\und{f}, p ] ; a_1 , \dots a_n) = \alpha ( [F(\und{f}), p] ; a_1 , \dots a_n) \end{equation}
for $p \in \PP (n)$, $c_i, t \in \CCC$ and $f_i: c_i \to t$  so $[\und{f}, p ] \in \PP_{\ovr{\CCC}}^r \ipt{t}{\und{c}}$, 
and $a_i \in (F^*A)_{c_i} = A_{F(c_i)}$. 
We will denote the left-adjoint of $F^*$ by $F_!$. 

So for any functor of orthogonal categories $F: \ovr{\CCC} \to \ovr{\DDD}$
we get the adjunction 
\begin{equation}\kern-7pt\xymatrix{
\FT ( \ovr{\CCC}, \PP, r_i)
\ar@<0.7ex>[r]^{ F_! }
& 
\FT ( \ovr{\DDD}, \PP, r_i)
\ar@<0.7ex>[l]^{ F^*}
}\!\!\!\!. \end{equation}
We will consider two specific orthogonal functors here, 
embeddings of full orthogonal subcategories and localizations, 
which are related to local-to-global constructions
and the time-slice axiom, respectively. 

We start with an embedding of a full orthogonal subcategory
$j: \ovr{\CCC} \to \ovr{\DDD}$. 
This means that as a functor $ \CCC \to \DDD$, $j$ is injective on objects
and is full and faithful, i.e.
\begin{equation} j : \CCC (c_1 , c_2 ) \to \DDD (j(c_1) , j( c_2)) \end{equation}
is bijective. Moreover, there are no more orthogonality relations on $\DDD$: 
$f_1 \perp_\CCC f_2$ if and only if $j(f_1) \perp_\DDD j(f_2)$. 

If we think of $\ovr{\DDD}$ as a space-time category, 
$\ovr{\CCC}$ will typically be a subcategory of particularly nice space-times. 
For example, we can consider $\Loc_\diamond \subseteq \Loc$, 
the subcategory of space-times whose underlying manifold is diffeomorphic to $\mR^n$. 
We can then ask if a theory $(A , \alpha)$ on $\Loc$
is determined by its behaviour on $\Loc_\diamond$, 
i.e. if it satisfies {\it descent} with respect to $\Loc_\diamond$. 
On the other hand, starting with a theory on $\Loc_\diamond$
we might look for a {\it local-to-global construction}
that extends it to a theory on $\Loc$. 
To this end, we have the following result. 

\begin{proposition}\label{jlocal}
If $j: \ovr{\CCC} \to \ovr{\DDD}$ is a full orthogonal subcategory embedding, 
$\FT ( \ovr{\CCC}, \PP, r_i)$ is a full coreflective subcategory of $\FT ( \ovr{\DDD}, \PP, r_i)$
through the adjunction
\begin{equation}\kern-7pt\xymatrix{
\FT ( \ovr{\CCC}, \PP, r_i)
\ar@<0.7ex>[r]^{ j_! }
& 
\FT ( \ovr{\DDD}, \PP, r_i)
\ar@<0.7ex>[l]^{ j^*}
}\!\!\!\!\!. \end{equation}
In other words, if $(A,\alpha) \in \FT ( \ovr{\CCC}, \PP, r_i)$
then $j^* j_! (A,\alpha) \cong (A, \alpha)$
via the unit of the adjunction. 
\end{proposition}

On the other hand, we call theories $(A,\alpha) \in \FT ( \ovr{\DDD}, \PP, r_i)$ {\it $j$-local}
if we have an isomorphism
\begin{equation}j_!  j^* (A, \alpha) \cong (A,\alpha)\end{equation} 
via the counit, 
i.e. if $(A, \alpha)$ satisfies descent with respect to $j: \ovr{\CCC} \to \ovr{\DDD}$. 
We write $\FT ( \ovr{\DDD}, \PP, r_i)^{j\text{-loc}}$
for the $j$-local theories in $\FT ( \ovr{\DDD}, \PP, r_i)$.
With Proposition \ref{jlocal} we immediately find that 
 \begin{equation}\xymatrix{
\FT ( \ovr{\CCC}, \PP, r_i)
\ar@<0.7ex>[r]^-{ j_! }
& 
\FT ( \ovr{\DDD}, \PP, r_i)^{j\text{-loc}}
\ar@<0.7ex>[l]^-{ j^*}
}  \end{equation}
is an adjoint equivalence. 

We interpret these results as follows: 
$j^*(A,\alpha)$ is the restriction of a field theory 
$(A,\alpha)$ on $\ovr{\DDD}$ to the full subcategory $\ovr{\CCC}$. 
On the other hand, $j_!$ is a local-to-global construction: 
it takes a theory on $\ovr{\CCC}$
and extends it to a theory on $\ovr{\DDD}$. 
Proposition $\ref{jlocal}$ then tells us that 
taking a theory on $\ovr{\CCC}$, extending it to $\ovr{\DDD}$ 
and then restricting it back to $\ovr{\CCC}$ doesn't change the theory, 
as we would expect from any reasonable procedure to extend a theory. 
Theories on $\ovr{\DDD}$ that are determined by the extension of their restriction 
are the $j$-local theories; 
these are exactly the theories that live in the essential image of $j_!$. 
As was shown in \cite{Benini:2017fnn}, $j_!$ is a generalization of 
Fredenhagen's universal algebra construction, \cite{Fredenhagen:1989pw,Fredenhagen:1993tx,Fredenhagen:1992yz,lang2014universal}. 

Next we turn to localization. 
Start with an orthogonal category $\ovr{\CCC} = (\CCC , \perp)$ 
together with a distinguished set of morphisms $W \subseteq \Mor(\CCC)$. 
We form the {\it localization} of $\CCC$ at $W$, $\CCC [W^{-1}]$ 
by formally inverting all morphisms in $W$. 
This construction comes with a localization functor $L: \CCC \to \CCC [W^{-1}]$.  
This in turn lets us define the push-forward relation $L_* (\perp)$ on $ \CCC [W^{-1}]$, 
which is the orthogonality relation generated by all $(L(f_1), L(f_2))$ where $f_1 \perp f_2$. 

So we have the {\it orthogonal localization} of $\ovr{\CCC}$ at $W$, 
$\ovr{\CCC[W^{-1}]} = (\CCC[W^{-1}] , L_*(\perp))$, 
and a functor of orthogonal categories $L: \ovr{\CCC} \to \ovr{\CCC[W^{-1}]}$. 
If $\ovr{\CCC}$ is a space-time category, $W$ will be the Cauchy morphisms. 
As mentioned earlier, we are interested in orthogonal localization 
because it implements the time-slice property: 
field theories on $\ovr{\CCC[W^{-1}]}$ 
will exactly be field theories $(A, \alpha)$ on $\ovr{\CCC}$ 
such that if $f: c \xrightarrow{} c'$ is a Cauchy morphism, 
$Af$ is an isomorphism. 
We can now consider the adjunction arising from $L$ to relate 
theories on $\ovr{\CCC}$ and theories on $\ovr{\CCC[W^{-1}]}$. 

\begin{proposition}\label{timeslice}
If $L: \ovr{\CCC} \to \ovr{\CCC[W^{-1}]}$ is an orthogonal localization, 
$\FT ( \ovr{\CCC[W^{-1}]}, \PP, r_i)$ is a full reflective subcategory of $\FT ( \ovr{\CCC}, \PP, r_i)$
through the adjunction
\begin{equation}\kern-7pt\xymatrix{
\FT ( \ovr{\CCC}, \PP, r_i)
\ar@<0.7ex>[r]^-{ L_! }
& 
\FT ( \ovr{\CCC[W^{-1}]}, \PP, r_i)
\ar@<0.7ex>[l]^-{ L^*}
}\!\!\!\!\!. \end{equation}
So if $(A,\alpha) \in \FT ( \ovr{\CCC[W^{-1}]}, \PP, r_i)$, 
$L_!  L^* (A , \alpha) \cong (A,\alpha)$
via the counit of the adjunction. 
\end{proposition}

On the other hand, we call field theories $(A,\alpha) \in \FT ( \ovr{\CCC}, \PP, r_i)$ {\it $W$-constant} 
if the unit gives an isomorphism
\begin{equation} (A, \alpha) \cong  L^* L_!(A,\alpha) \end{equation} 
and we write $\FT ( \ovr{\CCC}, \PP, r_i)^{W\text{-const}}$
for the $W$-constant theories on $\ovr{\CCC}$. 
By Proposition \ref{timeslice}, 
\begin{equation}\kern-7pt\xymatrix{
\FT ( \ovr{\CCC}, \PP, r_i)^{W\text{-const}}
\ar@<0.7ex>[r]^-{ L_! }
& 
\FT ( \ovr{\CCC[W^{-1}]}, \PP, r_i)
\ar@<0.7ex>[l]^-{ L^*}
} \end{equation}
is an adjoint equivalence. 

$L^*$ takes a theory on $\ovr{\CCC[W^{-1}]}$ 
and forgets that it satisfies the time-slice axiom. 
Conversely, $L_!$ is a time-slicification functor, 
taking a general theory on $\ovr{\CCC}$ and then generating one that satisfies the axiom. 
Proposition \ref{timeslice} implies that 
forgetting that a theory satisfies the time-slice axiom
and then generating a theory that satisfies it returns an isomorphic theory. 
So our time-slicification functor does not change theories already satisfying the axiom, 
as one would require from such a functor.

\subsection{Change of operad adjunctions}

As we saw, a map of bipointed operads 
\begin{equation}\phi: (\PP, r_i) \to (\QQ, s_i)\end{equation}
gives rise to a map of colored operads
\begin{equation} (\id_{\CCC_0} , \phi_{\ovr{\CCC}} ) :
 (\CCC_0 , \PP_{\ovr{\CCC}}^r )  \to (\CCC_0 , \QQ_{\ovr{\CCC}}^s )  \end{equation}
which in turn defines an adjunction
\begin{equation}\kern-7pt\xymatrix{
\FT ( \ovr{\CCC}, \PP, r_i)
\ar@<0.7ex>[r]^-{ (\id_{\CCC_0} , \phi_{\ovr{\CCC}} )_! }
& 
\FT ( \ovr{\CCC}, \QQ, s_i)
\ar@<0.7ex>[l]^-{  (\id_{\CCC_0} , \phi_{\ovr{\CCC}} )^*}
}\!\!\!\!\!. \end{equation}

We can again cast the pull-back $(\id_{\CCC_0} , \phi_{\ovr{\CCC}} )^*$ 
in a more concrete form: using that
\begin{equation}\FT ( \ovr{\CCC}, \QQ, s_i) \subseteq 
\big( \Alg (\QQ) \big)^\CCC\end{equation}
we find that
\begin{equation} (\id_{\CCC_0} , \phi_{\ovr{\CCC}} )^* 
= (\phi^*)_* : \FT ( \ovr{\CCC}, \QQ, s_i) \to \FT ( \ovr{\CCC}, \PP, r_i)\end{equation}
is the push-forward along the pull-back of $\phi$. 
Explicitly, 
\begin{equation} ((\phi^*)_* A)_c = A_c\end{equation}
and 
\begin{equation} ((\phi^*)_* \alpha) ( [\und{f} , p] ; a_1 , \dots a_n) 
= \alpha ( [\und{f} , \phi(p) ] ; a_1, \dots a_n )  \end{equation}
for $p \in \PP (n)$, $c_i, t \in \CCC$ and $f_i: c_i \to t$  
so $[\und{f}, p ] \in \PP_{\ovr{\CCC}}^r \ipt{t}{\und{c}}$, 
and $a_i \in A_{c_i} $. 

We will write $ (\phi^*)^!$ for the left adjoint
$(\id_{\CCC_0} , \phi_{\ovr{\CCC}} )_!$ of $ (\phi^*)_* $. 
If we consider the adjunction 
\begin{equation}\kern-7pt\xymatrix{
\Alg (\PP)
\ar@<0.7ex>[r]^-{ \phi_! }
& 
\Alg (\QQ)
\ar@<0.7ex>[l]^-{  \phi^*}
}  \end{equation}
from $\phi: \PP \to \QQ$
we would like to also have $(\phi^*)^! = (\phi_!)_*$. 
This is true in some cases, but not in general. 

So for a map of bipointed operads $\phi: (\PP, r_i) \to (\QQ, s_i)$
we have an adjunction
\begin{equation}\kern-7pt\xymatrix{
\FT ( \ovr{\CCC}, \PP, r_i)
\ar@<0.7ex>[r]^-{ (\phi^*)^! }
& 
\FT ( \ovr{\CCC}, \QQ, s_i)
\ar@<0.7ex>[l]^-{   (\phi^*)_* }
}\!\!\!\!\!. \end{equation}
Before we consider a specific map of bipointed operads, 
we want to know how this adjunction interacts 
with the constructions in the previous Section \ref{changecolor}. 

\begin{proposition}\label{preservepropo}
Let $\phi: (\PP, r_i) \to (\QQ, s_i)$ be a map of bipointed operads, 
$j: \ovr{\DDD} \to \ovr{\CCC}$ a full subcategory embedding
(note the reversal of names of orthogonal categories compared to earlier), 
and $L: \ovr{\CCC} \to \ovr{\CCC[W^{-1}]}$ an orthogonal localization. 
Then, for the adjunction 
\begin{equation}\kern-7pt\xymatrix{
\FT ( \ovr{\CCC}, \PP, r_i)
\ar@<0.7ex>[r]^-{ (\phi^*)^! }
& 
\FT ( \ovr{\CCC}, \QQ, s_i)
\ar@<0.7ex>[l]^-{   (\phi^*)_* }
}  \end{equation}
we have the following: 
\begin{enumerate}[i)]
\item $(\phi^*)_*$ preserves $W$-constant field theories; 
\item $(\phi^*)^!$ preserves $j$-local field theories; 
\item if $(\phi^*)^! = (\phi_!)_*: \FT ( \ovr{\EEE}, \PP, r_i) \to \FT ( \ovr{\EEE}, \QQ, s_i)$
for both $\ovr{\EEE} = \ovr{\CCC}$ and $\ovr{\EEE} = \ovr{\CCC[W^{-1}]}$, 
$(\phi^*)^!$ also preserves $W$-constant field theories. 
\end{enumerate}
\end{proposition}

So in a change-of-operad adjunction, 
both $j$-locality and $W$-constancy are not automatically preserved 
by both sides of the adjunction.

\subsection{Linear quantization}\label{linqu}

We will now construct a quantization adjunction, 
relating quantum field theories to linear field theories. 
Recall from Section \ref{operadicFT} that $\QFT (\ovr{\CCC}) = \FT ( \ovr{\CCC}, \As, \mu - \mu^{\text{op}} = 0)$ 
and $\LFT (\ovr{\CCC}) = \FT ( \ovr{\CCC}, \uLie , [,] = 0)$. 

The canonical way to define a Lie structure on an associative algebra $(A, \cdot)$ by 
\begin{equation} [a_1 , a_2 ] = a_1 \cdot a_2 - a_2 \cdot a_1\end{equation}
translates into a map of bipointed operads in the following way. 
On the level of operads, we have 
\begin{equation}
\begin{aligned} 
 \phi : \uLie & \to  \As  \\
 \begin{tikzpicture}[baseline=-6]
\draw (0,0.25) -- (0,0);
\draw (0,0) -- (-0.2,-0.2) node [below=-1pt] {\tiny{1}};
\draw (0,0) -- (0.2,-0.2) node [below=-1pt] {\tiny{2}};
\end{tikzpicture} & \mapsto  \begin{tikzpicture}[baseline=-6]
\draw (0,0.25) -- (0,0);
\draw (0,0) -- (-0.2,-0.2) node [below=-1pt] {\tiny{1}};
\draw (0,0) -- (0.2,-0.2) node [below=-1pt] {\tiny{2}};
\end{tikzpicture} - \begin{tikzpicture}[baseline=-6]
\draw (0,0.25) -- (0,0);
\draw (0,0) -- (-0.2,-0.2) node [below=-1pt] {\tiny{2}};
\draw (0,0) -- (0.2,-0.2) node [below=-1pt] {\tiny{1}};
\end{tikzpicture} \\
 \begin{tikzpicture}
\draw (0,0.25) -- (0,-0.1);
\draw (0,-0.1) circle (1.5pt) [fill=white]; 
\end{tikzpicture} & \mapsto \begin{tikzpicture}
\draw (0,0.25) -- (0,-0.1);
\draw (0,-0.1) circle (1.5pt) [fill=white]; 
\end{tikzpicture} 
\end{aligned}
\end{equation}
or in other words, $[,] \mapsto \mu - \mu^{\text{op}}$
where we recall that $\mu^{\text{op}}$ is the opposite multiplication. 
This map is well defined as can be easily checked, 
and it is consistent with the relations: 
\begin{equation} \phi \Big( \begin{tikzpicture}[baseline=-6]
\draw (0,0.25) -- (0,0);
\draw (0,0) -- (-0.2,-0.2) node [below=-1pt] {\tiny{1}};
\draw (0,0) -- (0.2,-0.2) node [below=-1pt] {\tiny{2}};
\end{tikzpicture} \Big) = \begin{tikzpicture}[baseline=-6]
\draw (0,0.25) -- (0,0);
\draw (0,0) -- (-0.2,-0.2) node [below=-1pt] {\tiny{1}};
\draw (0,0) -- (0.2,-0.2) node [below=-1pt] {\tiny{2}};
\end{tikzpicture} - \begin{tikzpicture}[baseline=-6]
\draw (0,0.25) -- (0,0);
\draw (0,0) -- (-0.2,-0.2) node [below=-1pt] {\tiny{2}};
\draw (0,0) -- (0.2,-0.2) node [below=-1pt] {\tiny{1}};
\end{tikzpicture} ~ , \quad 
\phi ( 0 ) = 0~. \end{equation}
So 
\begin{equation}\phi : ( \uLie , [,] = 0)  \to (\As, \mu - \mu^{\text{op}} = 0)  \end{equation}
is a morphism of bipointed operads
and we have an adjunction
\begin{equation}\kern-7pt\xymatrix{
\LFT (\ovr{\CCC}) 
\ar@<0.7ex>[r]^-{ (\phi^*)^! }
& 
\QFT (\ovr{\CCC})
\ar@<0.7ex>[l]^-{   (\phi^*)_* }
}  \end{equation}
for any orthogonal category $\ovr{\CCC}$. 

It turns out that $(\phi^*)^! = (\phi_!)_*$ 
so both the left and right adjoints are push-forwards along the uncolored adjunction
\begin{equation} \kern-7pt\xymatrix{
\Alg (\uLie )
\ar@<0.7ex>[r]^-{ \phi_! }
&
\Alg (\As)
\ar@<0.7ex>[l]^-{   \phi^* }
}\!\!\!\!\!. \end{equation}
The right adjoint $\phi^*$ is the map mentioned above: 
\begin{equation} \phi^* (A, \mu, e) = (A, [,] = \mu - \mu^{\text{op}}, e) . \end{equation}
It is well known that taking the universal enveloping algebra of a Lie algebra
is a left adjoint of this operation on general associative algebras. 
Our model for the left adjoint $\phi_!$ is the unital version hereof, 
identifying the unit of the tensor algebra $e_\otimes$ with our Lie algebra unit $e$. 

So defining $\UU = (\phi^*)_*$ and $\QQQ = (\phi_!)_*$ we have the adjunction
\begin{equation}\kern-7pt\xymatrix{
\LFT (\ovr{\CCC}) 
\ar@<0.7ex>[r]^-{ \QQQ }
& 
\QFT (\ovr{\CCC})
\ar@<0.7ex>[l]^-{ \UU }
}\!\!\!\!\!. \end{equation}
Note that with Proposition \ref{preservepropo},  
$\QQQ$ preserves both $j$-locality and $W$-constancy. 
We call $\QQQ$ the {\it linear quantization functor}. 

We justify the name we gave $\QQQ$ as follows. 
For a classical linear field theory, we have a vector space $V$ of linear observables, 
together with a symplectic form $\omega: V \otimes V \to \mR$. 
Canonical quantization of linear theories is a functor 
\begin{equation}\CCR: \Symp \to \Alg(\As)\end{equation}
that takes a classical theory $(V, \omega)$
and produces the associative algebra 
\begin{equation} T^\otimes V / I_{CCR} \end{equation}
where $T^\otimes V$ is the tensor algebra of $V$ 
and $I_{CCR}$ is the ideal generated by the relation
\begin{equation} v_1 \otimes v_2 - v_2 \otimes v_1 = i \omega(v_1,v_2) \oone~.  \end{equation}

Note that because $\omega: V \otimes V \to \mR$ is a 2-to-0 operation 
(and therefore not an $n$-to-1 operation)
$\Symp$ is not a category of algebras over an operad. 
So $\CCR$ does not arise as part of an adjunction from an operad map, 
and we cannot use the results found above directly. 
However, we can split up $\CCR$ and study a part of it. 

For any symplectic vector space $(V, \omega)$,
we can construct its {\it Heisenberg Lie algebra},
$\Heis (V, \omega) = (V \oplus i \mR , [,]_\omega)$, 
where 
\begin{equation} [ v_1 \oplus x_1 , v_2 \oplus x_2 ]_\omega = 0 \oplus i \omega(v_1 , v_2)~. \end{equation}
$\Heis (V, \omega)$ is a unital Lie algebra with $i \in i \mR$ the unit, 
so this gives rise to a functor 
\begin{equation} \Heis: \Symp \to \Alg (\uLie) \end{equation}
and we have
\begin{equation} \CCR = \phi_! \circ \Heis~. \end{equation}

A classical linear field theory now is a functor $\VV \in \Symp^\CCC$
such that for $f_1 \perp f_2$, 
$\omega \big( \VV(f_1) (-) , \VV (f_2) (-) \big) = 0$. 
If we then write $\CCR_*$ for the canonical quantization functor on these theories, 
\begin{equation} \CCR_* = \QQQ \circ \Heis_*\end{equation} 
explaining that $\QQQ = (\phi_!)_*$ is (half of) the linear quantization functor.

\section{Homotopy field theory and Quillen adjunctions}\label{homotopy}

In this section we take some first steps 
to refine the results from the previous sections to a model categorical setting.
Homotopy AQFT has earlier been studied in \cite{Benini:2018oeh} and \cite{Yau:2018dnm}. 
We will mostly give a sketch of our results, again referring to \cite{Bruinsma:2018knq} for details. 

In gauge theory, we work in a setting with higher structures, e.g. $\MMM = \Ch(k)$. 
Algebras of observables are differential graded algebras in the BRST/BV formalism, 
see \cite{Hollands:2007zg,Fredenhagen:2011an,Fredenhagen:2011mq} for an AQFT treatment hereof. 
Crucially $\Ch(k)$ is a {\it model category} \cite{hovey2007model}: 
it has a broader notion of equality than isomorphism called {\it weak equivalence}, 
which means that objects that are not isomorphic can still be equivalent. 
It also comes with two special classes of maps, {\it fibrations} and {\it cofibrations}. 
In $\Ch(k)$, the role of weak equivalences is played by quasi-isomorphisms: 
maps between complexes that are isomorphisms in homology. 

In general, functors do not preserve weak equivalences, 
which would lead to inconsistencies if we think of two weakly equivalent objects as being the same. 
In some cases this can be fixed. 
The usual procedure is as follows \cite{dwyer1995homotopy}. 
A {\it Quillen adjunction} is an adjunction
\begin{equation}\kern-7pt \xymatrix{
\CCC
\ar@<0.7ex>[r]^-{ F }
&
\DDD
\ar@<0.7ex>[l]^-{ G }
}  \end{equation}
such that $F$ preserves cofibrations, and $G$ preserves fibrations. 
We then introduce endofunctors $Q: \CCC \to \CCC $ and $R: \DDD \to \DDD$ 
such that $Qc$ is a cofibrant object for any $c \in \CCC$ and $Rd$ is a fibrant object for any $d \in \DDD$, 
together with natural weak equivalences $q: Q \to \id$ and $r: \id \to R$. 
We call $Q$ and $R$ {\it (co)fibrant replacement functors}. 
Then define $\mL F = F Q$ and $\mR G = G R$; 
these functors preserve weak equivalences. 
$\mL F$ and $\mR G$ are called {\it derived} functors for $F$ and $G$, respectively. 
We note that such derived functors are unique up to weak equivalence. 

For the rest of this section, let $\MMM = \Ch(k)$
with $k$ a field of characteristic zero. 
In this case, if $\PP \in \Op_\CC ( \Ch (k))$, 
$\Alg ( \PP )$ has a model structure 
where the weak equivalences are the quasi-isomorphisms on each color 
(i.e. $\kappa: A \to B$ such that $A_c \to B_c$ is a quasi-isomorphism for all $c \in \CC$)
\cite{Hinich:9702015,Hinich:1311.4130}. 
Moreover, in the context of field theories, 
if $F: \ovr{\CCC} \to \ovr{\DDD}$ is a morphism of orthogonal categories 
and $\phi: (\PP, r_i) \to (\QQ, s_i)$ is a morphism of uncolored bipointed operads 
then 
\begin{equation}\kern-7pt \xymatrix{
\FT ( \ovr{\CCC} , \PP, r_i )
\ar@<0.7ex>[r]^-{ (\phi_F)_! }
&
\FT ( \ovr{\DDD} , \QQ , s_i )
\ar@<0.7ex>[l]^-{ (\phi_F)^* }
}  \end{equation}
is a Quillen adjunction. 
One pleasant feature of the model structure on $\Alg ( \PP )$ 
is that all objects are fibrant, and therefore we can choose $R = \id$
when deriving functors on algebras of operads. 

From the preceding paragraphs we see that the linear quantization adjunction is a Quillen adjunction, 
and therefore there exists a derived linear quantization adjunction
\begin{equation}\kern-7pt \xymatrix{
\LFT (\ovr{\CCC}) 
\ar@<0.7ex>[r]^-{\mL \QQQ }
& 
\QFT (\ovr{\CCC})
\ar@<0.7ex>[l]^-{ \UU }
}\!\!\!\!\!. \end{equation}
In theory, this means that we have a functor to quantize linear gauge theories.
However, we need a workable model for the cofibrant replacement functor
to work with this construction in practice.

With these definitions, we can also recast 
our notions of $j$-locality and $W$-constancy to a model categorical setting. 
Note that since we have $R = \id$ we suppress any mention of $R$ and $r$. 
For a full subcategory embedding $j: \ovr{\CCC} \to \ovr{\DDD}$
we call a theory $(A, \alpha) \in \FT (\ovr{\DDD} , \PP , r_i)$
{\it homotopy $j$-local} if the derived counit gives a weak equivalence
\begin{equation} j_! Q j^* (A, \alpha) \simeq (A, \alpha)~. \end{equation}
For any theory $(A, \alpha) \in \FT ( \ovr{\CCC} , \PP , r_i)$, 
$\mL j_! (A, \alpha)$ is homotopy $j$-local. 
Proposition \ref{preservepropo} generalizes to this case: 
if $\phi : \PP^r \to \QQ^s$ is a morphism of bipointed operads, 
$\mL (\phi^*)^!$ preserves homotopy $j$-local field theories. 

For an orthogonal localization $L : \ovr{\CCC} \to \ovr{\CCC [ W^{-1} ] }$
we call a theory $(A , \alpha) \in \FT ( \ovr{\CCC} , \PP , r_i)$
{\it homotopy $W$-constant} if the derived unit gives a weak equivalence
\begin{equation} Q (A, \alpha) \simeq L^* L_! Q (A, \alpha)~. \end{equation}
The results about $W$-constancy do not translate as easily 
to the model categorical framework as those on $j$-locality. 
Extra assumptions are necessary to show that
$L^*$ maps to $W$-constant theories
and that $\mL (\phi^*)^!$ preserves $W$-constant field theories 
(including once more that $(\phi^*)^! = (\phi_!)^*$). 
We again refer for \cite{Bruinsma:2018knq} for details.

\section{Conclusion and outlook}\label{outlook}

In these proceedings we outlined an operadic way 
to formulate algebraic field theories on an orthogonal category. 
We saw that a field theory of type $(\PP , r_i)$ on $\ovr{\CCC}$ 
is an algebra over the $\CCC_0$-colored operad $\PP_{\ovr{\CCC}}^r$. 
Using this construction 
we were able to define local-to-global constructions and time-slicification, 
and we found a quantization functor for linear field theories. 
We then started a treatment of these ideas
in the context of model categories.  

In the future we hope to use these techniques 
to develop a suitable framework for constructing models 
of linear gauge theory in algebraic quantum field theory. 
In particular we want to formulate linear quantum Yang--Mills
and Chern--Simons theories in this setting. 
To do this several technical hurdles still need to be crossed. 

For one, a cofibrant replacement functor $Q$ as described in Section \ref{homotopy} always exists, 
but a general construction is typically very cumbersome. 
So one challenge is to find a suitable small enough model for 
the derived linear quantization functor $\mL \QQQ$. 

Of course, such an $\mL \QQQ$ is only a part of the story. 
For our construction we would also need a homotopically meaningful way 
to move from simple geometric data 
(e.g. a space of fields with an action functional)
to the category of linear field theories defined above
(i.e. field theories of type $(\uLie, [,] = 0)$). 
As in \cite{Gwilliam:2016mgk}, this would probably require
us to leave the framework of model categories.

\bibliography{allbibtex}

\bibliographystyle{prop2015}

\end{document}